%% file: reentrant.17.tex
\documentclass{jfm}

\usepackage{natbib}
\bibpunct{(}{)}{;}{a}{,}{,}
\usepackage{amsmath,amssymb,graphicx,afterpage,here}
\usepackage{afterpage,subfigure,epsfig,psfig,rotating,minitoc}
\usepackage[english,activeacute]{babel}
\usepackage{rangecite}
\usepackage{ae}
\usepackage{dcolumn}
\newcommand{\be}{\begin{equation}}
\newcommand{\ee}{\end{equation}}
\newcommand{\bea}{\begin{eqnarray}}
\newcommand{\eea}{\end{eqnarray}}

\newcommand{\noi}{\noindent}

\newcommand{\pd}{\partial}

\newcommand{\HR}[2]{{}{{ #2}}}

\newcommand{\LB}[1]{\label{#1}}

\input jfmheader.tex

\title{Reentrant Hexagons in non-Boussinesq Convection\\}

\author[]{Santiago Madruga\dag, Hermann Riecke\dag, and  Werner Pesch\ddag}
\affiliation{{\small \dag Department of Engineering Sciences and Applied Mathematics,
Northwestern University, Evanston, IL 60208, USA\\
\ddag Physikalisches Institut, Universit\"at Bayreuth, D-95440 Bayreuth, Germany}}

\date{\today}
\pagerange{\pageref{firstpage}--\pageref{lastpage}}

\begin{document}

\maketitle

\typeout{sigue, sigue ....!Sputnik ....}

\begin{abstract}

While non-Boussinesq hexagonal convection patterns are well
known to be stable close to threshold (i.e. for Rayleigh numbers
$R \approx R_c$), it has often been assumed that they are always
unstable to rolls already for slightly higher Rayleigh numbers.
Using the {\em incompressible} Navier-Stokes equations for
parameters corresponding to water as a working fluid, we perform
full numerical stability analyses of hexagons in the strongly
nonlinear regime ($\epsilon\equiv R-R_c/R_c=\mathcal{O}(1)$). We
find `reentrant' behavior of the hexagons, i.e. as $\epsilon$ is
increased they can lose and {\it regain} stability. This can
occur for values of $\epsilon$ as low as $\epsilon=0.2$. We
identify two factors contributing to the reentrance: i) the
hexagons can make contact with a hexagon attractor that has been
identified recently in the nonlinear regime even in Boussinesq
convection (\cite{AsSt96,ClBu96}) and ii) the non-Boussinesq
effects increase with $\epsilon$. Using direct simulations for
circular containers we show that the reentrant hexagons can
prevail even for side-wall conditions that favor convection in
the form of the competing stable rolls. For sufficiently strong
non-Boussinesq  effects  hexagons become stable even over the
whole $\epsilon$-range considered, $0 \le \epsilon \le 1.5$. 

\end{abstract}

\tableofcontents



\section{Introduction}
 
Rayleigh-B\'enard convection has served as an excellent paradigm
for studying systems that spontaneously form spatial or
spatio-temporal patterns. In recent years exciting results have
been obtained for the stability and dynamics of structures that
are connected with roll convection. To mention are, in
particular, the spiral-defect chaos obtained in convection of
fluids with low Prandtl number and domain chaos driven by the
K\"uppers-Lortz instability in rotating systems. For a recent
review see \cite{BoPe00}. In most of these investigations great
care has been taken to keep the experimental systems close to the
regime in which the Oberbeck-Boussinesq approximation is valid by
minimizing the dependence of the fluid parameter on the
temperature in order to avoid the appearance of cellular or
hexagonal structures. 

It has long been recognized that variations of the fluid
parameters with the temperature, i.e. non-Oberbeck-Boussinesq
(NOB) effects, break an up-down symmetry  and therefore introduce
otherwise prohibited mode interactions. The best-studied case is
the resonant triad interaction among the three fundamental
Fourier modes whose wavevectors form a hexagonal pattern. It
renders the primary bifurcation to the hexagons transcritical
with the consequence that the hexagons are preferred over rolls
in the immediate vicinity of onset (\cite{Bu67}). According to
leading-order weakly nonlinear analysis, hexagons typically
become unstable to rolls further above threshold where the
amplitudes are larger and the resonant-triad interaction loses
significance compared to interactions involving four modes
(\cite{Pa60,SeSt62,Se65,Bu67,PaEl67,DaSe68}). This scenario of a
transition from hexagons to rolls has been confirmed in quite a
number of experimental investigations
(\cite{SoDo70,DuBe78,Ri78,WaAh81,BoBr91,PaPe92}) and it has
commonly been assumed that in convection hexagonal patterns are
confined to the regime close to onset. Presumably due to this
assumption, hexagonal convection in
strongly nonlinear non-Boussinesq convection has not been
investigated much.

Recently, however, two experiments showed that even in the
strongly nonlinear regime stable hexagonal convection patterns
can be observed. In one experiment hexagons were observed at
relatively high Rayleigh numbers ($\epsilon \equiv (R-R_c)/R_c
\approx 3.5$) under conditions in which the Oberbeck-Boussinesq
(OB) approximation was quite well satisfied (\cite{AsSt96}). In
that case hexagons with up-flow and those with down-flow in the
center are equivalent and both were observed to coexist in
adjacent domains. A subsequent numerical stability analysis
confirmed the existence of such stable OB-hexagons 
(\cite{ClBu96,BuCl99a}). They are characterized by a pronounced
spatial concentration of the flow in the off-center regions. To
resolve the strong spatial variations of the hexagons in the
plane modes with relatively high wavenumbers, which are neglected
in the familiar amplitude equations, need to be retained
(\cite{BuCl99}). 

In the other experiment, which used $SF_6$ near the thermodynamic
critical point as the working fluid, it was found that the
hexagons that arise in the immediate vicinity of the onset of
convection can become unstable as the Rayleigh number is
increased and then {\em restabilize} again at higher Rayleigh
numbers, $\epsilon = {\mathcal O}(1)$ (\cite{RoSt02}).  This
restabilization was termed {\it 'reentrance'}. As the
non-Boussinesq effects were increased the intermediate
$\epsilon$-range over which rolls were the preferred planform
shrank and eventually hexagons were found to dominate rolls from
onset all the way to $\epsilon = {\mathcal O}(1)$.  Since the
reentrant hexagons have been observed even for moderate
NOB-effects  the restabilization at larger $\epsilon$ has been
attributed by the authors to the high compressibility of $SF_6$
near its critical point (\cite{RoSt02}).

In previous numerical stability calculations of {\it rotating}
non-Boussinesq convection in water it was found that hexagons
can be linearly stable over the whole range  $0\le \epsilon \le
1$ (\cite{YoRi03b}). In the strongly nonlinear regime $\epsilon
= {\mathcal O}(1)$ a chaotic state (`whirling chaos') was
obtained in which individual hexagonal cells oscillate or
rotate, often inducing the nucleation of additional cells. Even
in the presence of lateral walls, which typically induce the
nucleation of rolls, this hexagon-based spatio-temporally
chaotic state was found to persist. In that investigation the
mechanism that is responsible for the linear stability of the
hexagons over a fairly wide range in $\epsilon$  was, however,
not understood. 

In the present paper we identify easily accessible parameter
regimes of non-Boussinesq convection in which strongly nonlinear
hexagons are linearly stable.  We point out two mechanisms that
contribute to their stability. First, the mechanism that
stabilizes OB-hexagons at larger Rayleigh numbers also enhances
the stability of the NOB-hexagons. Second, with increasing
Rayleigh number the temperature difference across the fluid layer
increases and with it the strength of the NOB-effects. A simple
weakly nonlinear model shows that this effect alone can be
sufficient to lead to a restabilization of the hexagons already
at relatively low values of $\epsilon$, which experimentally
leads to reentrant hexagons. Together, these two mechanism can
lead to a restabilization for values of $\epsilon$ as low as
$\epsilon = 0.2$. Since our computations are based on water as a
working fluid (cf. \cite{YoRi03b}), which is essentially
incompressible in the investigated regime, it is clear that high
compressibility is not a necessary  condition to obtain reentrant
hexagons.

The paper is organized as follows. In
Sec.\ref{sec:basicequations} we briefly review the
basic equations that we use, pointing out in which way our
computations focus  on weakly non-Boussinesq, but strongly
nonlinear convection. In Sec.\ref{sec:stability} we present our
results for the linear stability regimes of hexagonal and roll
patterns. The two mechanisms leading to restabilization are
discussed in Sec.\ref{sec:mechanism}. Direct simulations of the
temporal evolution for 
boundary conditions mimicking circular containers are discussed
in Sec.\ref{sec:simulations}. Conclusions follow in
Sec.\ref{sec:conclusions}.

\section{Basic Equations \LB{sec:basicequations} }

We consider a horizontal fluid layer of thickness $d$, density
$\rho$, kinematic viscosity $\nu$, heat conductivity $\lambda$,
and specific heat $c_p$.  The layer is infinite in the
horizontal extension and is limited in the vertical direction by
two horizontal, rigid plates with high thermal conductivity. The
system is heated from below (at  temperature $T_1$) and cooled
from above (at temperature $T_2 < T_1$).   The governing
equations expressing the balance of momentum, mass, and energy
are (\cite{Ch61})
\bea 
\pd_t (\rho u_i)+\pd_j (\rho u_j u_i)&=&-\pd_ip -\rho g\delta_{i3}+\pd_j \left(\nu\rho\,(\pd_iu_j+\pd_ju_i)\right)\\
\pd_t\rho+\pd_j(\rho u_j)&=&0  \LB{e:conti}\\
 \pd_tT+u_j\pd_jT &=&\frac{1}{\rho c_p}\pd_j(\lambda \pd_jT). \LB{e:heat00}
\eea
Here $\vec{u}=(u_1,u_2,u_3)$ is the fluid velocity, $T$ the
temperature, $p$ the pressure, $g$ the acceleration of gravity,
and  $\delta_{ij}$  is the Kronecker delta. Summation over
repeated indices is implied. The origin of a Cartesian
coordinate system with the $z$-axis perpendicular to the
horizontal plates is fixed in  the middle of the layer.  As
usually, viscous heating and volume viscosity effects can safely
be neglected.

Realistic rigid boundary conditions are taken at the two
boundaries for the velocity,
\bea
\vec{u}=0 \mbox{ at } z=\pm\frac{d}{2}
\eea
\noi and fixed values for the temperature 
\bea
T&=T_1\equiv&T_0+\frac{\Delta T}{2}  \mbox{ at } z=-\frac{d}{2}, \LB{e:bcTbot}\\ 
T&=T_2\equiv&T_0-\frac{\Delta T}{2}  \mbox{ at } z=+\frac{d}{2}. \LB{e:bcTtop}
\eea 
Here $T_0 = (T_1 + T_2)/2$ denotes the mean temperature and
$\Delta T=(T_1 -T_2) > 0$ is the temperature difference across the
layer. We assume an experimental procedure in which $T_0$ is
kept constant  while the main control parameter $\Delta T$ is
varied.

We focus in this work on weakly non-Boussinesq convection, so we
keep the temperature dependence of the various fluid properties
to leading order, and expand them about the mean temperature
$T_0$ in line with Busse's convention (\cite{Bu67})
\bea
\frac{\rho(T)}{\rho_0}&=&1-
\bar{\gamma}_0 \frac{T-T_0}{T_s}(1+\bar{\gamma}_1\frac{T-T_0}{T_s})+... \label{e:NBrho}\\
\frac{\nu(T)}{\nu_0} &=& 1+\bar{\gamma}_2 \frac{T-T_0}{T_s}+...\\
\frac{\lambda(T)}{\lambda_0} &=& 1+\bar{\gamma}_3\frac{T-T_0}{T_s}+...\label{e:NBlambda}\\
\frac{c_p(T)}{c_{p0}}&=&1+\bar{\gamma}_4\frac{T-T_0}{T_s}+...\label{e:NBcp}
\eea
where $\rho_0$, $\nu_0$, $\lambda_0$, and $c_{p0}$ denote the
values of the respective quantities at the mean temperature
$T_0$. The dots denote higher-order terms to be neglected.
Introducing the thermal diffusivity
$\kappa_0=\lambda_0/\rho_0c_p$, the scaling temperature  $T_s =
\nu_0 \kappa_0/(\alpha_0 g d^3)$ is used to define the
nondimensionalized    slopes of the density, viscosity, heat
conductivity, and heat capacity at $T_0$  in terms of the
$\bar{\gamma}_i$ For instance, the usual heat expansion
coefficient at $T = T_0$ is given by
$\alpha_0=\bar{\gamma}_0/T_s $. Beyond the Boussinesq
approximation also the curvature of $\rho (T) $ at $T_0$, which
is proportional to $\bar{\gamma}_0 \bar{\gamma}_1/T_s^2$, comes
into play. 

To make the governing equations and boundary conditions
dimensionless,  the following  scales are selected: for the
length $d$, for the time $d^{2}/\kappa_0$,  for the pressure
$\rho_0\nu_0 \kappa_0/d^{2}$, and for the temperature $T_s$.
This gives rise to two dimensionless quantities:  the Prandtl
number $Pr=\nu_0/\kappa_0$, and  the Rayleigh number $R= \Delta
T/T_s = \alpha_0 \Delta T g d^3/(\nu_0 \kappa_0)$.   Thus, we
use in the following the dimensionless temperatures $\hat T
=T/T_s$ and $\hat{T}_0=T_0/T_s$,  heat conductivity $\hat
\lambda =\lambda/\lambda_0$, density  $\hat \rho =\rho/\rho_0$,
kinematic viscosity $\hat \nu =\nu/\nu_0$, and specific heat
$\hat c_p =c_p/c_{p0}$. Finally, we write the equations  in
terms of the dimensionless momenta $v_i=\rho u_i d /\rho_0
\kappa_0$  instead of the velocities. In the following we omit
the hats again for simplicity. 

Since the fluid velocities are small compared to the sound velocity
we make the anelastic approximation (\cite{Go69}) \typeout{(see also
\cite{Bu89})} and neglect the time derivative in the continuity
equation (\ref{e:conti}). This simplifies the computation
considerably since it reduces the number of evolution equations.
Furthermore, $v_i$ becomes a solenoidal field, which can be
represented in the standard poloidal-toroidal decomposition by two
velocity potentials (\cite{Bu89}) automatically enforcing the
mass conservation.

The (dimensionless) conduction solution ($\vec v = 0$) of (\ref{e:heat00}) with
(\ref{e:bcTbot},\ref{e:bcTtop},\ref{e:NBrho},\ref{e:NBlambda},\ref{e:NBcp})
is given by
\bea
T_{cond}=T_0+
R\left(-z-\frac{\gamma_3}{2}(z^2-\frac{1}{4})+{\mathcal O}(\gamma_3^2)\right).\LB{e:cond}
\eea
We rewrite the temperature $T$ in terms of the deviation $\Theta$ from the 
conductive profile neglecting its ${\mathcal O}(\gamma_3^2)$-contribution,
\bea 
\Theta=T-T_{cond}=T-T_0-R\left(-z-\frac{\gamma_3}{2}(z^2-\frac{1}{4})\right). \LB{e:Tdev}
\eea

We then obtain as the final dimensionless equations
\bea
\frac{1}{Pr}\left(\pd_tv_i+v_j\pd_j\left(\frac{v_i}{\rho}\right)\right)&=&-\pd_i p 
+\delta_{i3}\left(1+\gamma_1(-2 z+\frac{\Theta}{R})\right)\Theta+\nonumber \\
&&+\pd_j\left[\nu\rho\left(\pd_i(\frac{v_j}{\rho})+\pd_j(\frac{v_i}{\rho})\right)\right]
\LB{e:v}\\
\pd_jv_j&=&0, \LB{e:cont}\\
\pd_t\Theta+\frac{v_j}{\rho}\pd_j\Theta
& =&\frac{1}{\rho
c_p}\pd_j(\lambda\pd_j\Theta)-\gamma_3\pd_z\Theta-R\frac{v_z}{\rho}(1+\gamma_3z).\LB{e:T}
\eea
The dimensionless boundary conditions are
\bea
\vec{v}(x,y,z,t)=\Theta(x,y,z,t)=0  \mbox{ at } z= \pm \frac{1}{2}.\LB{e:bc}
\eea
The nondimensionalized  fluid parameters (\ref{e:NBrho})-(\ref{e:NBcp}) read now:
\bea
\rho(\Theta)&=&1-\gamma_0(-z+\frac{\Theta}{R}),\LB{e:rhoTh}\\
\nu(\Theta)&=& 1+\gamma_2(-z+\frac{\Theta}{R}),\LB{e:nuTh}\\
\lambda(\Theta)&=&1+\gamma_3(-z+\frac{\Theta}{R}),\LB{e:lambdaTh}\\
c_p(\Theta)&=&1+\gamma_4(-z+\frac{\Theta}{R}).\LB{e:cpTh}
\eea
where we have
introduced the quantities $\gamma_i \equiv \bar {\gamma}_i
R = \bar{\gamma}_i \Delta T / T_s $, which  can be expressed as
follows: 
\bea \label{eq:gam}
\gamma_i(\Delta T) = \gamma_i^{c}\, \left(1+\frac{\Delta T-\Delta
T_c}{\Delta T_c}\right)=\gamma_i^{c}\, \left(\frac{R}{R_c} \right).
\eea
Here $\gamma_i^{c} = \bar{\gamma}_i \Delta T_c/T_s$
denotes the non-Boussinesq coefficients evaluated at the onset of
convection, $T = T_c$, as used in \cite{Bu67}.  In our case the
$\gamma_i^{c} = \bar{\gamma}_i \Delta T_c / T_s $ can be well
approximated  using the Boussinesq value for the critical
temperature difference, $\Delta T_c = 1708 \,T_s$. In
principle, the non-Boussinesq effects induce small corrections
through a shift in the threshold, which are easily
numerically computed, if necessary. Note that in contrast
to the $\bar{\gamma}_i$  the $\gamma_i(\Delta T)$ are linear in
the main control parameter  $\Delta T$ and therefore have to be
adjusted in the nonlinear regime.

We consider the non-Boussinesq effects to be weak and keep in all
material properties only the leading-order temperature dependence
beyond the Boussinesq approximation. Therefore the
$\gamma_1$-term appears explicitly in  (\ref{e:v}), while in all
other terms  it would constitute only a quadratic correction just
like the terms omitted in (\ref{e:NBrho})-(\ref{e:NBcp}).
Correspondingly, we expand the denominators in
(\ref{e:v},\ref{e:T}) that contain material properties  to
leading order in the $\gamma_i$. In analogy to \cite{Bu67}, we
further omit non-Boussinesq terms that contain cubic
nonlinearities in the amplitudes $v_i$ or $\Theta$, as they arise
from the expansion of the advection terms $v_j
\partial_j(v_i/\rho)$ and $(v_j/\rho)\partial_j \Theta$ when the
temperature-dependence of the density is taken into account. 
Since we  will be considering Rayleigh numbers up to twice the
critical value, which implies enhanced non-Boussinesq effects,
these approximations may lead to quantitative differences 
compared to the fully non-Boussinesq system, even though the
temperature-dependence of the material properties themselves may
quite well be described by a linear (or quadratic in the case of
the density) approximation.   Unfortunately we are not aware of
rigorous solutions of Eqs (2) that would allow
quantitative  tests of our approximations scheme for finite
$\gamma_i$.

In the weakly nonlinear regime hexagon and roll patterns are
described by the amplitudes $A_i$ of the three dominant Fourier
modes associated with the wavevectors ${\bf q_1} = q(1,0),\, {\bf
q_2} = q/2 (-1, \sqrt{3}),\, {\bf q_3} = q/2 (-1, -\sqrt{3})$.  
Roll solutions correspond to $A_1\ne 0, \,A_2 = A_3 = 0$ and
hexagons to $A_1 = A_2 = A_3\ne 0$, while for mixed hexagon
solutions one has $ A_1\ne A_{2,3} \ne 0$. The amplitudes satisfy
the well-known coupled amplitude equations (e.g. \cite{CrHo93}), 
\bea   
\partial_t A_1 &=& \epsilon A_1
-\delta\overline{A}_2\overline{A}_3 -g_1\,|A_{1}|^{2}
A_1-g_2(|A_{2}|^{2}+|A_{3}|^{2})A_1 \LB{ampli-redu}    
\eea 
with the equations for $A_{2,3}$ obtained from (\ref{ampli-redu})
by cyclic permutation. The calculation of the coefficients
$\delta$, $g_i$ involves vertical averages over certain products
of the critical eigenvector components, which are obtained from
a linearization of (\ref{e:v},\ref{e:T}).  

The quadratic coefficient $\delta$ arises from the NOB-effects.
To leading order in the $\gamma_i^{c}$ it has been calculated first
by \cite{Bu67} and is proportional to Busse's parameter $Q$,
which is conventionally used as a  measure for the strength of the
NOB-effects. It is defined as 
\bea
Q = \sum_{i=0}^{4}\gamma_i^{c} {\cal P}_i,\LB{e:busseq}
\eea
\noi 
where the ${\cal P}_i$ are certain linear functions of
$Pr^{-1}$, which can be found in the review article by 
\cite{BoPe00}\footnote{The expressions given on p. 742 of
\cite{BoPe00} use the symbol $\cal P$ instead of $Q$ and
correct a small error in Busse's calculation of ${\mathcal
P}_3$.}.  Busse's parameter $Q$  characterizes the breaking of
the up-down symmetry, which renders at most one of the two
possible types of hexagons stable. Gases have a positive value
of $Q$ and exhibit hexagons with down flow in the center
($g$-hexagons), whereas liquids have negative $Q$ and show
hexagons with up flow ($l$-hexagons). In the weakly nonlinear
approach the cubic coefficients $g_i$ in (\ref{ampli-redu}) are
evaluated with $\gamma_i^{c} = 0$, which is consistent with the
assumption of small $\gamma_i^{c}$, $\gamma_i^{c}={\mathcal
O}(A_i)$. In the case of finite $\gamma_i^{c}$ the calculation
can be refined along the lines presented by \cite{PlPe99}.

The stability of weakly nonlinear roll and hexagon patterns  is
determined by a linear stability analysis of the various
solutions of Eq. (\ref{ampli-redu}).   The resulting bifurcation
diagram  is  sketched schematically in Fig.\ref{fig.sub}  for
the case $g_2>g_1$ (cf. \cite{Bu67}). The hexagons arise
unstably from the conductive state in a transcritical
bifurcation and become stable through a saddle-node bifurcation.
When the heating, i.e. $\epsilon$, is increased they become 
unstable  in a transcritical bifurcation at $\epsilon_H$
involving a mixed-mode  solution. The rolls in turn are unstable
at threshold and  become stable through a pitchfork bifurcation
at $\epsilon_R$, when the control parameter is increased.

\begin{center}
\begin{figure} 
\centering
\includegraphics[width=0.7\textwidth,angle=0]{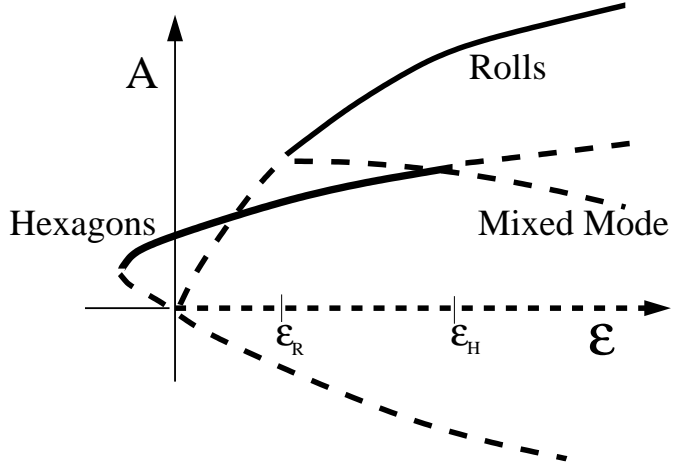}
\caption{Sketch of the bifurcation diagram for hexagons and
rolls in the weakly non-linear regime of NOB-convection. The
solid lines correspond to stable states and dashed lines to
unstable ones. The  unstable mixed mode destabilizes the hexagons
and stabilizes the rolls.  
\LB{fig.sub}} 
\end{figure} 
\end{center}

We focus in this paper on the stability properties of the
patterns in the strongly nonlinear regime. They are
determined by a Galerkin expansion of all fields in Eq.
(\ref{e:v},\ref{e:cont},\ref{e:T}) (see, for instance,
\cite{BuCl79a,Bu89}). Their dependence on the vertical
coordinate $z$ is captured by expanding them
in appropriate combinations of trigonometric and Chandrasekhar
functions in $z$  that satisfy the top and bottom boundary
conditions (\cite{Ch61,Bu67}). In most of the computations we
used $n_z=6$ modes for each field. With respect to the horizontal
coordinates in the lateral directions we use a Fourier expansions
on a hexagonal lattice.  The Fourier wave vectors ${\bf q}$ are
constructed as linear combinations  of the hexagonal basis
vectors ${\bf b}_1 =q(1,0)$ and  ${\bf b}_2 =q(1/2, \sqrt{3}/2)$
as  ${\bf q} = m {\bf b}_1 + n {\bf b}_2$ with the integers $m$
and $n$ in the range  $|m {\bf b}_1+n{\bf b}_2 | \le n_q q$.  The
largest wavenumber is then $n_q q$ and the number of Fourier
modes retained is given by $1+6\sum_{j=1}^{n_q}j$. Typically we
used $n_q =3 $. Thus, solving the PDEs
(\ref{e:v},\ref{e:cont},\ref{e:T}) is reduced to solving a system
of nonlinear ODEs in time for the Galerkin expansion coefficients. The standard
linear analysis of the ODEs yields the critical Rayleigh number
$R_c$ as well as the critical wavenumber $q_c$. Both depend on
the NOB-coefficients $\gamma_i^{c}$ which in turn depend on
$R_c$. Thus, in principle one obtains an implicit equation for
the $\gamma_i^{c}$. The shift in the critical Rayleigh number
away from the classical value $R_c=1708$ due to the NOB-effects
is, however, quite small (less than 1 percent) and therefore the
resulting change in the $\gamma_i^{c}$ is negligible. In this
paper we therefore choose the $\gamma_i^{c}$ corresponding to
$R_c=1708$. 

To investigate the nonlinear hexagon solutions, we start with
the standard weakly nonlinear analysis to determine the
coefficients of the coupled amplitude equations
(\ref{ampli-redu}). To obtain the fully nonlinear solutions we
need to solve the ODEs for the coefficients of the Galerkin 
expansion, which become a system of nonlinear algebraic
equations in our stationary case.  This is achieved with a
Newton solver for which the weakly nonlinear solutions serve as
convenient starting solutions. The solutions are tested for
amplitude stability by monitoring the growthrates of linear
perturbations of the expansion coefficients. Since the system is
spatially periodic (characterized by  Fouriermodes with
wavevectors $\bf q$) the possibility of side-band instabilities
with respect to modes with wavevectors  $\bf q \pm s $ has to be
considered as well. This is achieved by introducing Floquet
multipliers  $\exp (i {\bf s}\cdot (x,y))$ in the Fourier ansatz
for the linear perturbations of the Galerkin solutions. 

We also study the dynamics of complex patterns that arise from
instabilities of the periodic states. For that purpose we have
extended our previously
developed spectral code for the OB-equations
(\cite{Pe96,BoPe00}) to include the NOB-effects in
(\ref{e:v},\ref{e:cont},\ref{e:T}).  It employs the same vertical
modes as the Galerkin stability code but places the wave
vectors of the Fourier modes on a rectangular rather than a
hexagonal grid. To solve for the time dependence we have chosen
a fully implicit scheme for the linear terms, whereas the
nonlinear parts are treated explicitly (second-order
Adams-Bashforth method). The time step is typically taken to be
$t_v/500$, where $t_v$ is the vertical diffusion time. We have
tested that the stability regimes obtained from the Galerkin
analysis are consistent with the direct numerical simulations.

\section{Linear Stability of Hexagons  \LB{sec:stability} }

Instead of extensive parameter studies, we present in this
work  specific, interesting scenarios that should be
experimentally realizable. We focus our investigation on water,
which has a moderate Prandtl number and for which reentrant
hexagons should be readily accessible in convection cells
with  conventional layer thickness $d$ in a range of
temperatures close to room temperature.

\subsection{Amplitude Instabilities}

In our analysis, we first concentrate on spatially periodic
solutions with the wavenumber fixed at the critical wavenumber
and discuss their domains of existence and stability as a
function of the control parameter
$\epsilon=(R-R_c(\gamma_i^{c}))/R_c(\gamma_i^{c})$.  We have
chosen three different cells with thickness $d=1.5,\, 1.8,
\mbox{ and } 2.1\, mm$, respectively.  The case $d=1.8\,mm$ is
of particular interest since it has been studied in previous
experiments (\cite{PaPe92}).

Tab. \ref{t:water} gives  the non-Boussinesq coefficients and the
value of the non-Boussinesq parameter $Q$ at the onset of
convection for a representative range of mean temperatures $T_0$
in a fluid layer of thickness $d=1.8\,mm$\footnote{These values
were obtained with a code kindly provided by G. Ahlers.}.  As
indicated before, the $\gamma_i$ are linear in the temperature
difference $\Delta T$ (see \ref{eq:gam}) and therefore they
depend on $\epsilon$, 
\bea \gamma_i = \gamma_i^{c}\,\,
\left(1+\frac{R-R_c}{R_c}\right)=\gamma_i^{c} \,\,
(1+\epsilon). \LB{e:gameps} 
\eea  Note that with increasing mean
temperature  the critical temperature difference $\Delta T_c$
(given in the second column of Tab. \ref{t:water})  decreases.
Therefore the variation of the fluid properties across the layer
at the critical temperature and with it the coefficients
$\gamma_i^{c}$ also decrease  with increasing temperature. Since
we keep the mean temperature $T_0$ fixed when changing the
Rayleigh number, the accessible range in $\epsilon$ is limited by
the requirement that the temperature at the top plate, $T(z=d/2)
= T_0 - \Delta T /2$, be above freezing. It turns  out that the
full range $0 \le \epsilon \le 1$ is then only accessible for
average temperatures $T_0$ above a certain  temperature  $T^f_0$,
which is  $20^o C$ for $d = 1.8 \, mm$.

\begin{table}
\begin{tabular}{ccccccccc}\hline 
$T_0$ [$^oC$]& $\Delta T_c$ [$^oC$] & $Pr$ & $\gamma_0^{c}$ &
$\gamma_1^{c}$ & $\gamma_2^{c} $ & $\gamma_3^{c}$ &
$\gamma_4^{c}$ & $Q$ \\ \hline 
20 &20.63 & 6.93 & 0.0042&   0.5693 &  -0.5186 &  0.0649 &  -0.0049 & -4.612  \\ \hline   
25 &15.16 & 6.10  & 0.0038  & 0.2952 & -0.3370 & 0.0434  & -0.0022 & -2.489    \\ \hline
28 &12.94 & 5.68 & 0.0036  & 0.2122 & -0.2725 & 0.0352  & -0.0013 & -1.837    \\ \hline
32 &10.74 & 5.18 & 0.0034  & 0.1440 & -0.2126 & 0.0273  & -0.0005 & -1.292     \\ \hline
36 &9.12 & 4.76  &  0.0032 &  0.1023&  -0.1709& 0.0216  &  0.0001 & -0.954    \\ \hline    
40 & 7.87 & 4.38 & 0.0030 & 0.0755 &  -0.1405 & 0.0173 & 0.0004 &  -0.731  \\ \hline 
50 & 5.75 & 3.62 & 0.0026 & 0.0400 &  -0.0926 & 0.0104 & 0.0007 &  -0.428  \\ \hline 
60 & 4.42 & 3.05 & 0.0023 & 0.0245 &  -0.0654 & 0.0064 & 0.0006 &  -0.287  \\ \hline       
\end{tabular}  
\caption{Values of the  Prandtl number $Pr$, non-Boussinesq coefficients
$\gamma_i^{c}$,  and Busse's parameter $Q$ for water at the
onset of convection as a function of the mean temperature and
the temperature difference. The liquid layer has a depth of
$d=1.8\,mm$.  \LB{t:water} } 
\end{table} 

\begin{center}
\begin{figure}
\centering
\includegraphics[width=0.7\textwidth,angle=270]{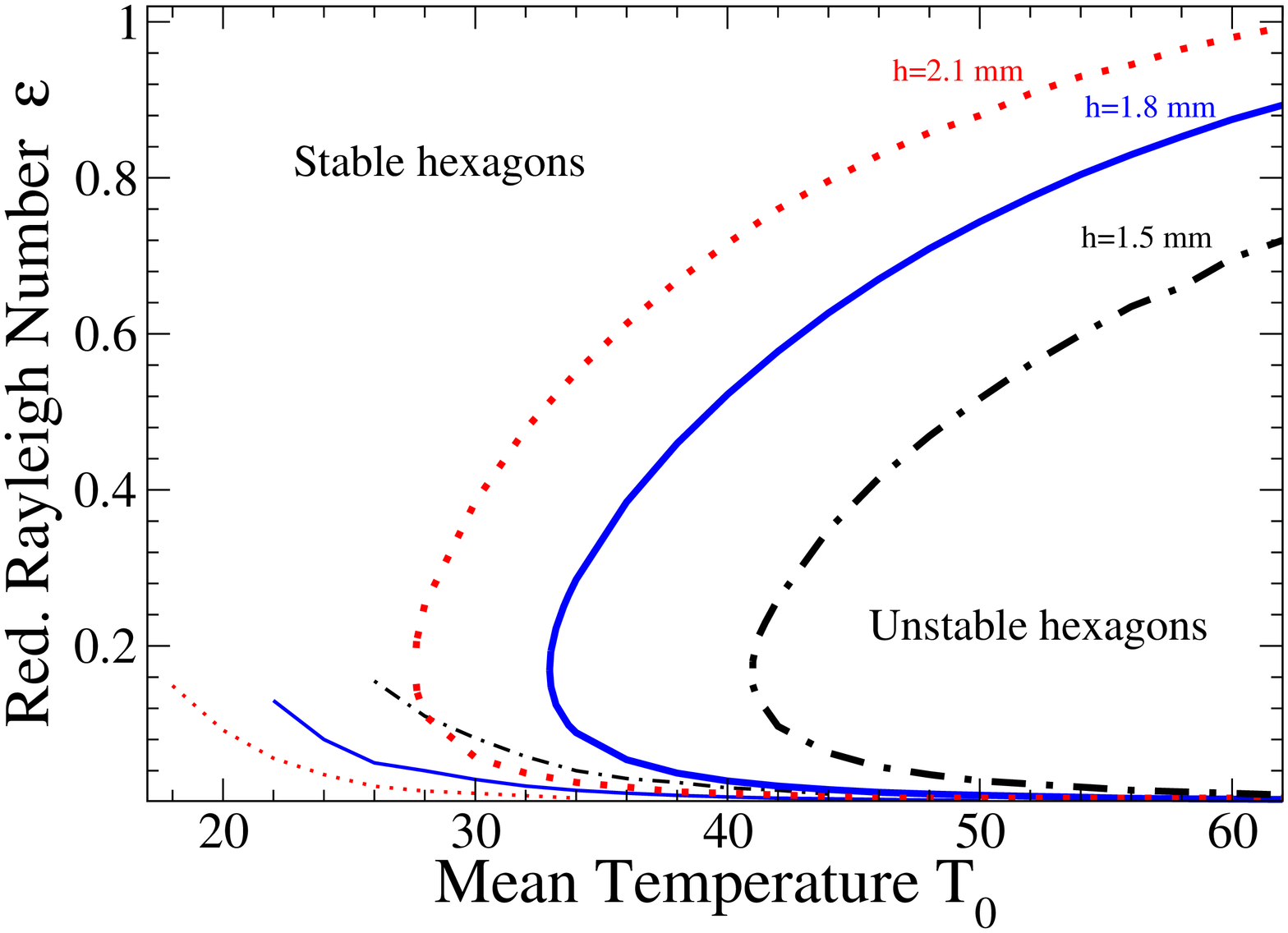}
\caption{Stability regions for water with respect to amplitude 
perturbations for three fluid depths: $d=2.1\, mm$ (dotted
lines), $d=1.8\, mm$ (full lines), $d=1.5\, mm$ (dot-dashed line).
Thick curves: stability boundaries for hexagons. Thin curves: stability
boundaries for rolls. For a given depth, rolls are stable above the thin 
line, and hexagons unstable in the inner region of the thick line. 
Stability limits obtained for the critical wavenumber $q_c$. \LB{ampli-water}}
\end{figure}
\end{center}

Using the Galerkin method, we extend the weakly nonlinear result
sketched in Fig.\ref{fig.sub} (\cite{Bu67}) to the strongly
nonlinear regime. Fig.\ref{ampli-water} shows the resulting
stability limits for hexagons and for rolls. As predicted by
weakly nonlinear theory, the hexagons are linearly stable with
respect to amplitude perturbations for very small $\epsilon$ with
respect amplitude perturbations. For not too small values of the
mean temperature $T_0$ and layer thickness $d$ they become
unstable as the control parameter is increased. This instability
corresponds to the transcritical bifurcation at $\epsilon_H$ in
Fig.\ref{fig.sub}.  For the convection cells investigated here,
the hexagon patterns then undergo a second steady bifurcation as
the control parameter is increased further and become stable
again.  As the mean temperature is decreased or the layer
thickness is decreased the critical heating and with it the
non-Boussinesq effects increase. This shifts the point of
restabilization to lower $\epsilon$ and the lower stability limit
to higher $\epsilon$, decreasing the $\epsilon$-range over which
the hexagons are unstable, until the two limits merge at a
temperature $T_m$. For $T_0<T_m$ the hexagons are
amplitude-stable over the whole range of $\epsilon$ considered
($0 \le \epsilon \le 1$).  

\begin{center} 
\begin{figure}
\begin{center} 
\subfigure[]{
\LB{water:bublea} 
\includegraphics[width=0.6\textwidth,angle=270]{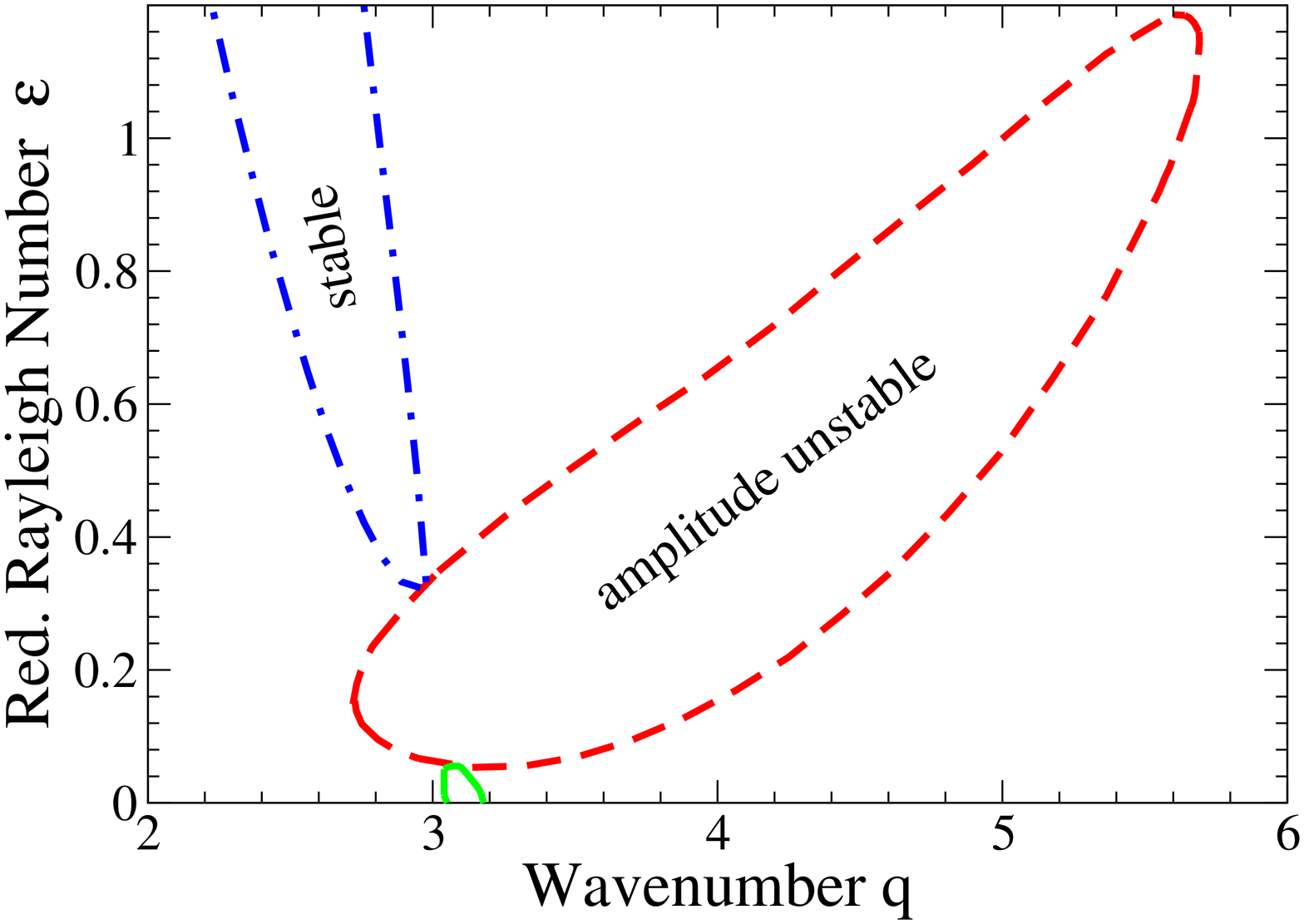}   
} 
\subfigure[]{
\LB{water:bubleb} 
\includegraphics[width=0.6\textwidth,angle=270]{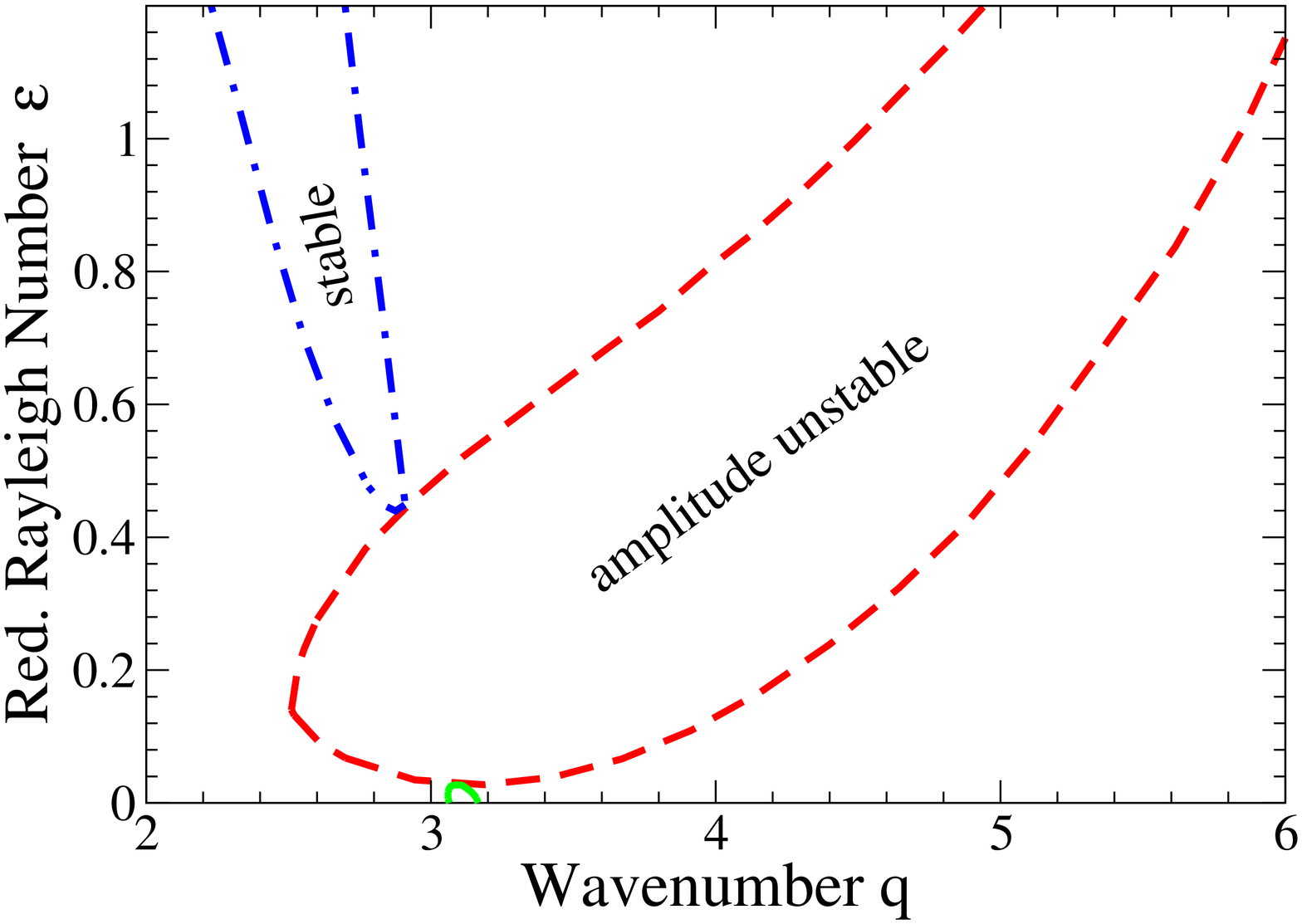}  
} 
\end{center} 
\caption{ 
 Stability regions for hexagons in water with respect
to amplitude  (dashed line) and side-band perturbations 
(dot-dashed line). The depth of the fluid layer is $d=1.8\,
mm$. Figure (a) $T_0=36\,^oC$, (b) $T_0=40 \,^oC$. Hexagons are
stable with respect to amplitude perturbations outside the
dashed-line region, and stable with respect to side-band
perturbation inside the dot-dashed line region.\LB{water:buble}}
\end{figure} 
\end{center} 

The restabilization of hexagons in the strongly nonlinear regime
has been observed by Roy and Steinberg in
$S\,F_6$ near the thermodynamical critical point, where it has
been termed `reentrance'  (\cite{RoSt02}). There it was argued that since the
non-Boussinesq effects in that system are not very large near
onset the reentrance is due to the large compressibility of the
fluid in this parameter regime. By assuming that the working
fluid is incompressible, which is an excellent approximation for
water, our computations show that high compressibility is not
needed for reentrance. The significance of the compressibility
for the occurrence of the reentrance has also been called into
question recently by \cite{Ah05} (see his footnote 59) based on
the work by \cite{OhOr04}.

We have also computed the stability of rolls with respect to
amplitude perturbations. The corresponding stability limits are
indicated  in Fig.\ref{ampli-water}  by  thin lines. Note that it
is not meaningful to extend these stability limits to lower
values of $T_0$ than shown in Fig.\ref{ampli-water} since then
the temperature at the top plate would be below the freezing
temperature of water. Below the thin curves  rolls are unstable, 
but they become linearly stable when $\epsilon$ is increased
beyond the lines. This stability limit corresponds to the
pitch-fork bifurcation at $\epsilon_R$ in Fig.\ref{fig.sub}. As
the non-Boussinesq effects become stronger the stabilization of
rolls is shifted to larger $\epsilon$. In contrast to the
hexagons, for the convection cells investigated here  the rolls
do not undergo a second bifurcation, which would destabilize
them, and remain amplitude-stable up to the largest values of
$\epsilon$ considered. For strong non-Boussinesq effects one has
therefore a very large range of parameters over which the
competing rolls and  hexagons are both linearly
amplitude-stable. 

\begin{center}
\begin{figure}
\centering
\includegraphics[width=0.62\textwidth,angle=270]{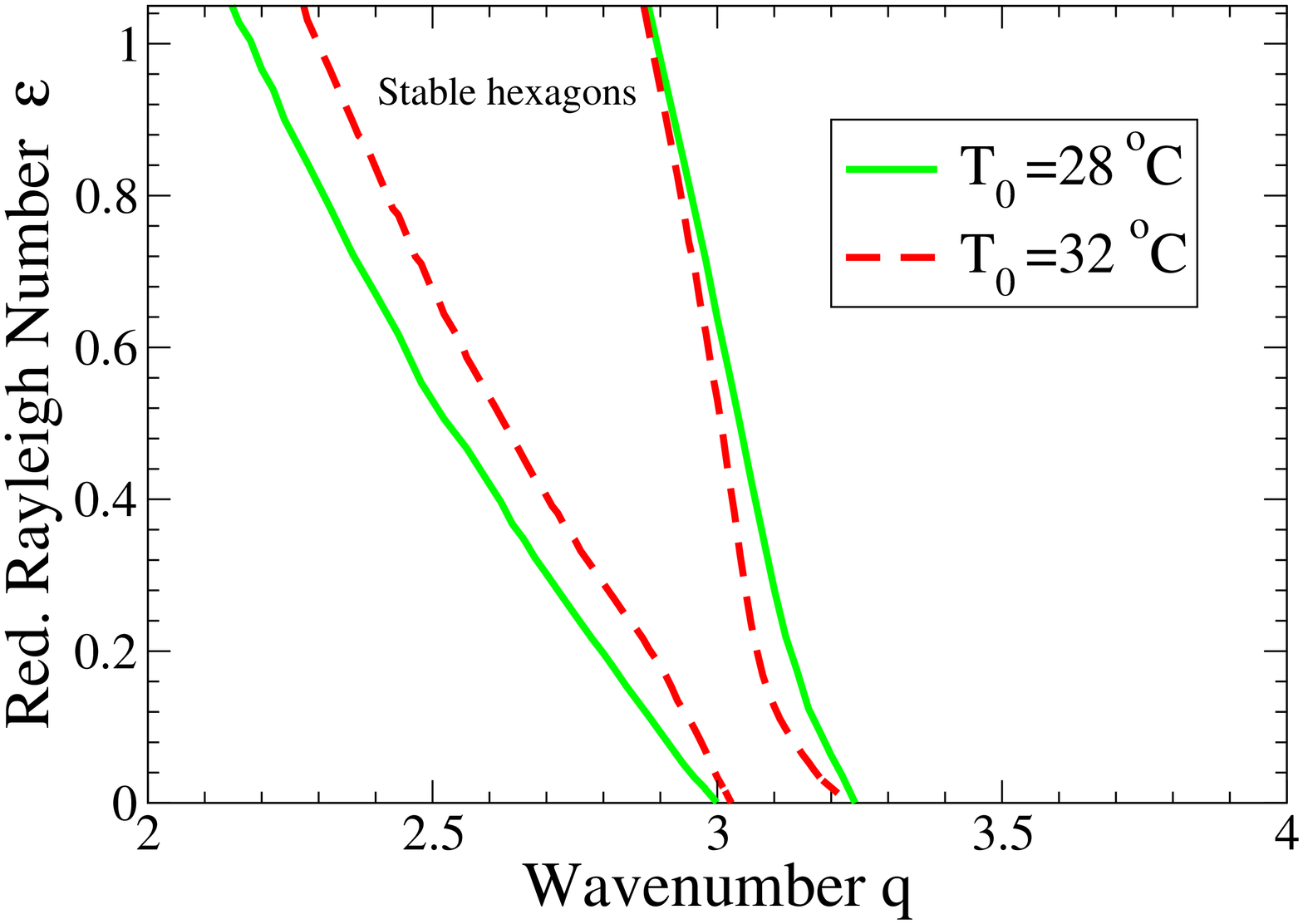}
\caption{
Stability regions for hexagons with respect to side-band pertubations, in water.
 The fluid depth is  $d=1.8\, mm$.
Solid line: $T_0=28^o C$. Dashed line: $T_0=32^o C$. \LB{to28-32}} 
\end{figure}
\end{center}

\HR{would it be worthwhile to add the side-band instabilities of
the Boussinesq case to one of the figures to make the connection
with Busse more explicit yet?}

\HR{let's take the limits in $\epsilon$ equal in figs.3 and 4}

The amplitude-stability limits of the hexagons and rolls depend, of
course, on their wavenumber. This is illustrated in Fig.
\ref{water:buble}, where we fix the mean temperature $T_0$ and
determine the stability limits of the hexagons as a function of
their wavenumber $q$. Interestingly,   for $T_0=36\,^o C$
(Fig.\ref{water:bublea}) the hexagons become more stable for
small and for large wavenumbers and the instability region forms
a bubble-like, closed curve, outside of which the hexagons are
stable with respect to amplitude perturbations. These bubbles
are hyper-surfaces in $q-\epsilon-T_0$-space. With decreasing
non-Boussinesq effects (increasing $T_0$)  the bubble grows and 
extends to larger values of $\epsilon$. Eventually, the upper
part of the bubble is shifted to $\epsilon$-values beyond the
range considered in this paper (Fig.\ref{water:bubleb} for
$T_0=40^oC$). 

\subsection{Side-Band Instabilities}

Using the Galerkin method, we have studied the stability of the
hexagons with respect to long- and short-wave perturbations as
shown in Fig.\ref{water:buble}. We find that over the whole
range $0\le \epsilon \le 1$ the only relevant side-band
perturbations are long-wave and steady, as is the case in the
weakly nonlinear regime. The long-wave perturbations involve
longitudinal and transverse phase modes, which can be described
by two coupled evolution equations (\cite{LaMe93,Ho95,EcPe98}).
The stability limits obtained from the Galerkin analysis are
shown in Fig.~\ref{water:buble}. In this parameter regime the
stability region consists of two disconnected domains,
reflecting the reentrant nature of the hexagons. The stability
domain near onset is very small and closes up as the amplitude
stability limit is reached. This behavior corresponds to that
obtained from the weakly nonlinear theory (\cite{LaMe93}). In
the reentrant regime the stable domain opens up again in an
analogous fashion  when the amplitude-stability limit is
surpassed. Note that the stability boundaries  lean toward lower
wavenumbers. Thus, stable reentrant hexagonal patterns are
expected to have wavenumbers below $q_c$. 

As the mean temperature is lowered the bubble of the amplitude
instability shrinks and eventually the bubble disappears
(Fig.\ref{to28-32}). The side-band stability limit reaches then
without interruption from the strongly nonlinear regime all the
way down to threshold (more precisely to the saddle-node
bifurcation of the hexagons). As the non-Boussinesq effects
become yet stronger the range of stable wavenumbers widens. 


\section{Origin of Reentrant Hexagons\LB{sec:mechanism} }

At first the appearance of stable reentrant non-Boussinesq
hexagons seems quite surprising, in particular, considering the
relatively small values of $\epsilon$ for which the
restabilization of the hexagons can occur. We have identified two
major factors that contribute to their appearance. One factor is
the fact that even in the Boussinesq case hexagons can be stable
for sufficiently large Rayleigh numbers (above $\epsilon \approx
1$). They have been observed in convection experiments using
$SF_6$ close to its thermodynamical critical point as a working
fluid where they nucleated in the cores of target and spiral
patterns (\cite{AsSt96}). A subsequent numerical stability
analysis confirmed the existence of stable OB-hexagons and
attributed their appearance to the formation of plumes
(\cite{ClBu96,BuCl99a}). The second factor contributing to the
reentrance of the non-Boussinsesq hexagons is the increase of the
non-Boussinesq effects with the Rayleigh number (cf.
(\ref{e:gameps})). 

Fig.\ref{NB-coeff} provides a quantitative assessment of the
importance of the two mechanisms contributing to the reentrance
of non-Boussinesq hexagons. Above the dashed-dotted line hexagons
with wavenumber $q_c$ become amplitude-stable in the Boussinesq
case $\gamma_i=0$. This transition line corresponds to the
stability limit of Boussinesq hexagons found earlier by
\cite{ClBu96} (see their Fig.4). The very slight dependence of
the stability limit on the mean temperature is due to the
variation of the Prandtl number with $T_0$, which decreases from
$Pr=5.4$ at $T_0=30\,^oC$ to $P=3.0$ at $T_0=60\,^oC$. With the
non-Boussinesq effects included, the stabilization of the
hexagons occurs at lower values of the control parameter. The
dashed line in Fig.\ref{NB-coeff} shows the resulting stability
limits when the $\epsilon$-dependence of the $\gamma_i$ is
neglected, $\gamma_i=\gamma_i^{c}$, while the solid line denotes
the stability limit with the dependence retained,
$\gamma_i=\gamma_i^{c}(1+\epsilon)$. Thus, even when the
non-Boussinesq effects are kept constant the stability limit
connected with the stabilization of Boussinesq hexagons merges
with the usual low-$\epsilon$ stability limit when the
non-Boussinesq effects become strong enough, i.e. at low
temperatures.

\begin{center}
\begin{figure}
\centering
\includegraphics[width=0.6\textwidth,angle=270]{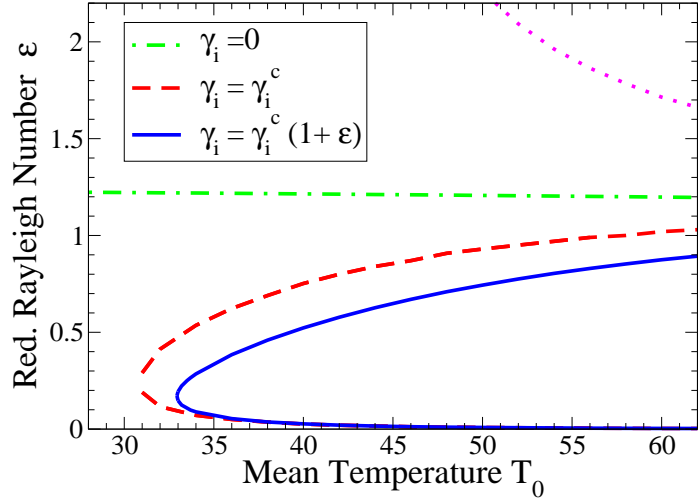}
\caption{Stability regions for hexagons with respect to
amplitude  perturbations in water for  $d=1.8\, mm$. Solid line:
stability limit when the dependence of $\gamma$ on the 
temperature is taken into account, $\gamma_i = \gamma_i
^{c}(1+\epsilon)$. Dashed line: stability limit when  the
$\gamma_i$ are fixed to their critical value, $\gamma_i=\gamma_i
^{c}$. OB-hexagons are stable above the dashed-dotted line,
$\gamma_i=0$. Down-hexagons are stable above the dotted line
for $\gamma_i = \gamma_i ^{c}(1+\epsilon)$.  \LB{NB-coeff}}
\end{figure}
\end{center}

Since in the Boussinesq-case up- and down-hexagons are
equivalent and become stable simultaneously it is to be expected
that for weak non-Boussinesq effects both types of hexagons can
become stable for large Rayleigh numbers, with the stabilization
occurring, however, at different values of the Rayleigh number.
For the up-hexagons, which in water are stable near onset, this
stabilization corresponds to a reentrance. The stabilization of
the down-hexagons is indicated in Fig.\ref{NB-coeff} by a dotted
line. As the non-Boussinesq effects become stronger the
down-hexagons require ever higher Rayleigh numbers for
stabilization. 
 
Even though the $\epsilon$-dependence of the non-Boussinesq
effects is not the central driving force for the reentrance, it
is instructive to discuss its effect on the hexagons briefly
within a weakly nonlinear framework. We consider as a minimal
model a slightly generalized version of Eq.(\ref{ampli-redu}) in
which the quadratic coupling coefficient grows linearly with
$\epsilon$,
\bea 
\partial_t A_1 &=& \xi^2 ({\bf n}_1\cdot \nabla)^2 A_1 +\epsilon
A_1 -
(\delta+\mu \epsilon) \overline{A}_2\overline{A}_3
-g_1\,|A_{1}|^{2}A_1\LB{e:amp}\\
&& -g_2(|A_{2}|^{2}+|A_{3}|^{2})A_1 
\nonumber
\eea 

Compared to (\ref{ampli-redu}) we also include a spatial gradient
term involving the normal derivative ${\bf n}_i\cdot \nabla$ with
${\bf n}_i={\bf q}_i/|{\bf q}_i|, i=1,2,3$. It allows long-wave
modulations of the amplitude that capture side-band
instabilities. Note that such a linear correction of the
quadratic resonance term has been considered previously in the
context of hexagonal patterns in ferrofluids exposed to a
magnetic field (\cite{FrEn01}). 

To get a complete picture of the impact of the
$\epsilon$-dependence of the non-Boussinesq effects within the
framework of the simple model (\ref{e:amp}) we introduce a
rescaled amplitude ${\mathcal A}=g_1A/(\delta+\mu \epsilon)$, a
rescaled time $\hat{t}=(\delta+\mu \epsilon)^2t/g_1 $, and a
rescaled space variable $\hat{\mathbf{r}}=(\delta + \mu
\epsilon){\mathbf r}/\xi \sqrt{g_1}$. Eq.(\ref{e:amp}) can then
be written as 

\begin{eqnarray}
\partial_{\hat{t}} {\mathcal A}_1=({\bf n}_1\cdot \hat{\nabla})^2 {\mathcal A}_1
+ \hat{\epsilon}{\mathcal A}_1 - \bar{{\mathcal A}}_2
\bar{{\mathcal A}}_3 - |\mathcal{A}_1|^2\mathcal{A}_1  
-\frac{g_2}{g_1}\left(|{\mathcal A}|_2^2+|{\mathcal
A}|_3^2\right){\mathcal A}_1. \LB{e:ampscal}
\end{eqnarray}
Here the rescaled control parameter is given by
\begin{equation}
\hat{\epsilon}=\frac{\epsilon g_1}{(\delta +\mu
\epsilon)^2},\LB{e:ehat}
\end{equation}
which implies that each value of $\hat{\epsilon}$ corresponds to
two values of $\epsilon$,  $\epsilon_{1,2}=(g_1-2\delta \mu
\hat{\epsilon} \pm \sqrt{g_1^2-4g_1\delta \mu
\hat{\epsilon}})/2\mu^2\hat{\epsilon}$.

\begin{center}
\begin{figure}
\centering
\includegraphics[width=0.6\textwidth,angle=0]{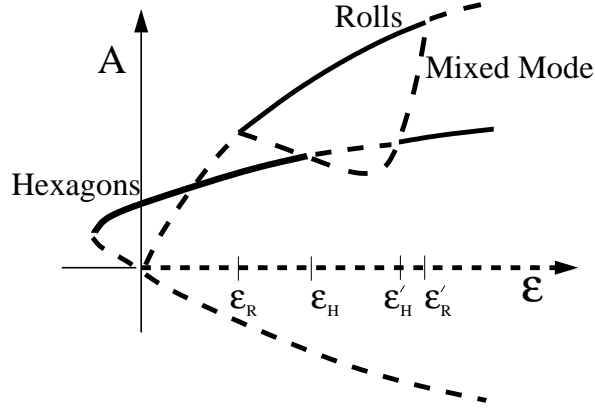}
\caption{Sketch of the bifurcation diagram obtained from
(\ref{e:amp}) (cf. Fig.\ref{f:quartic} below for $k=0$).}
\LB{f:bifreentrant}
\end{figure}
\end{center}

For the case $g_1>0$ and $\mu \delta >0$, which is of interest
here, a monotonic increase in $\epsilon$ is mapped into a
non-monotonic change of $\hat{\epsilon}$ with the range
$0\le\epsilon \le \epsilon_m\equiv \delta/\mu$ mapped onto $0\le
\hat{\epsilon} \le g_1/4\delta \mu$ and the range $\epsilon_m \le
\epsilon < \infty$ mapped in a reverse fashion onto the same
interval, $g_1/4\delta \mu \ge \hat{\epsilon} > 0$. Thus, with
increasing  $\epsilon$ the standard bifurcation diagram sketched
in Fig.\ref{fig.sub} is traversed towards the right up to
$\epsilon=\epsilon_m$. Note that the parameter $\epsilon$ in
Fig.\ref{fig.sub} plays the role of $\hat{\epsilon}$ in
(\ref{e:ampscal}). As $\epsilon$ is increased further
$\hat{\epsilon}$ decreases implying that the path through the
bifurcation diagram in Fig.\ref{fig.sub} is reversed. Thus, if
$\epsilon_m$ is not too small the same mixed-mode that is created
in the bifurcation  stabilizing the rolls and that destabilizes
the hexagons at a larger value of $\epsilon$ in a transcritical
bifurcation restabilizes the hexagons in a second transcritical
bifurcation and eventually disappears at the bifurcation that
destabilizes the rolls again. This scenario is shown in the
qualitative bifurcation diagram depicted in
Fig.\ref{f:bifreentrant}. For larger values of $\mu$,
\bea
\mu>\mu_M\equiv\frac{1}{2\delta}\,\frac{(g_1-g_2)^2}{2(2g_1+g_2)},
\eea
$\hat{\epsilon}$ does not reach the stability limit of the
hexagons and the hexagons remain linearly stable for all values
of $\epsilon$. Specifically, for $\mu=\mu_m$ the two
transcritical bifurcations at $\hat{\epsilon}_H$ and
$\hat{\epsilon}'_H$ coincide and the upper and lower stability
limits in Fig.\ref{NB-coeff} merge, eliminating the
amplitude-unstable regime of the hexagons.

\begin{center} \begin{figure} \centering
\includegraphics[width=0.6\textwidth,angle=0]{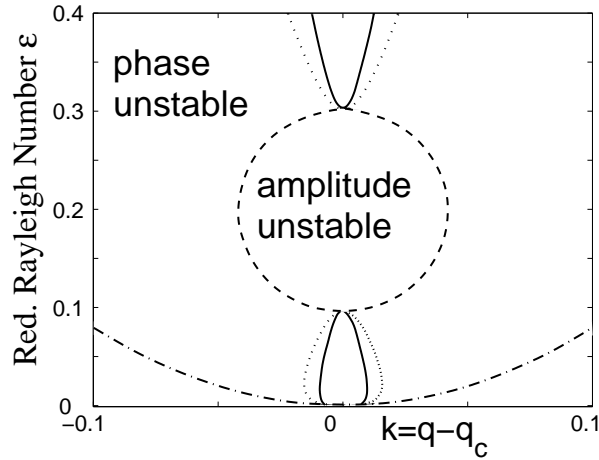}
\caption{Amplitude and side-band stability limits based on the
extended Ginzburg-Landau-equation model (\ref{e:amp}) with
$\xi=2.83$, $\delta=0.048$, $\mu=0.28$, $g_1=1$, and
$g_2=1.45$.  Shown are the neutral curve (dashed-dotted), the
long-wave side-band stability limit (solid and dotted), and the
amplitude stability limit (dashed).  Hexagons are stable inside
the solid lines.} \LB{f:quartic}  \end{figure}  \end{center}

Fig.~\ref{f:quartic} gives the side-band stability limits
obtained within the minimal model (\ref{e:amp}) for a typical
case. The solid lines denote the transverse long-wave phase
mode, while the dotted line marks the longitudinal phase mode.
The bifurcation diagram shown in Fig.\ref{f:bifreentrant}
corresponds to traversing this phase diagram at $q=q_c$. Note
that the destabilization of rolls at $\epsilon'_R$ is beyond the
range of $\epsilon$ shown in Fig.~\ref{f:quartic}. In principle,
either of the phase modes can lead to an instability. Within
eq.~(\ref{e:amp}), however, the longitudinal phase mode is relevant
only for an extremely small range of $\epsilon$ near threshold.
For small Prandtl numbers eq.~(\ref{e:amp}) would have to be
extended to include a mean flow. Then the longitudinal phase
mode can dominate the transverse mode over significant portions
of the stability limits (\cite{YoRi02,SeSc04}).

Thus, the minimal model (\ref{e:amp}) captures qualitatively the
restabilization of the hexagons, the merging of the two stability
limits as the non-Boussinesq effects are increased, and the main
features of the side-band instabilities obtained in the full
stability computations described in Sec.\ref{sec:stability}. We
have not found, however, the destabilization of rolls by a steady
mode as it is suggested by the minimal model (cf.
Fig.\ref{f:bifreentrant}). Instead, we find at quite large
$\epsilon$ an oscillatory instability.

The full numerical stability analysis presented in
Sec.\ref{sec:stability} displayed a tendency of the
side-band-stable regions to shift to lower wavenumbers as
$\epsilon$ is increased. In principle, this could be modeled
phenomenologically by retaining nonlinear gradient terms in the
minimal model (\cite{BrVe98,EcPe98,NuNe98}). We will not pursue
this here. Instead we point out that in experiments and in
numerical stability analyses the Boussinesq hexagons are found to
be stable with respect to side-band instabilities only for
wavenumbers noticeably below $q_c$ (\cite{AsSt96,ClBu96}).
Considering the significance of the mechanism underlying the
stability of the Boussinesq hexagons for the reentrance of the
non-Boussinesq hexagons, it is to be expected that the reduced
wavenumber is a characteristic feature of this mechanism of
reentrance. 

\section{Numerical Simulations\LB{sec:simulations}}

To make closer contact with the results that would be expected
in experimental investigations we perform also direct numerical
simulations of (\ref{e:v},\ref{e:T}). While our Galerkin
approach uses realistic boundary conditions at the top and
bottom plate it employs  periodic boundary in the lateral
directions. Thus, these computations are able to predict
instabilities that arise in the interior of the system, but they
do not capture phenomena associated with the lateral walls. In
most of the recent experiments circular containers have been
used and typically it has been found that the walls prefer
rolls, which are predominantly either perpendicular or parallel
to the wall depending on details of the side-wall conditions.
Thus, even very close to onset, where in the non-Boussinesq case
rolls are unstable to hexagons, in these experiments a narrow
ring of roll-like structures arises. 

A consequence of the wall-induced preference of rolls over
hexagons is the experimental observation that the transition
from hexagons to rolls does not occur at the Rayleigh number at
which the hexagons become linearly unstable, but already at
lower Rayleigh numbers (\cite{BoBr91}). The transition is
therefore more appropriately interpreted as arising from a
competition between rolls and hexagons, which are simultaneously
linearly stable in that regime. The lowest-order Ginzburg-Landau
equation (\ref{e:amp}) is variational and therefore, within this
framework, the transition is expected to occur when the energy of
the rolls becomes lower than that of the hexagons. The full
Navier-Stokes equations are not variational and therefore further
away from threshold it is more appropriate to discuss the
transition in terms of the invasion of one state into the other
with the velocity of the front separating the two states going
through zero at the transition point. 
 
\begin{center}
\begin{figure}
\centering
\includegraphics[width=0.9\textwidth,angle=0]{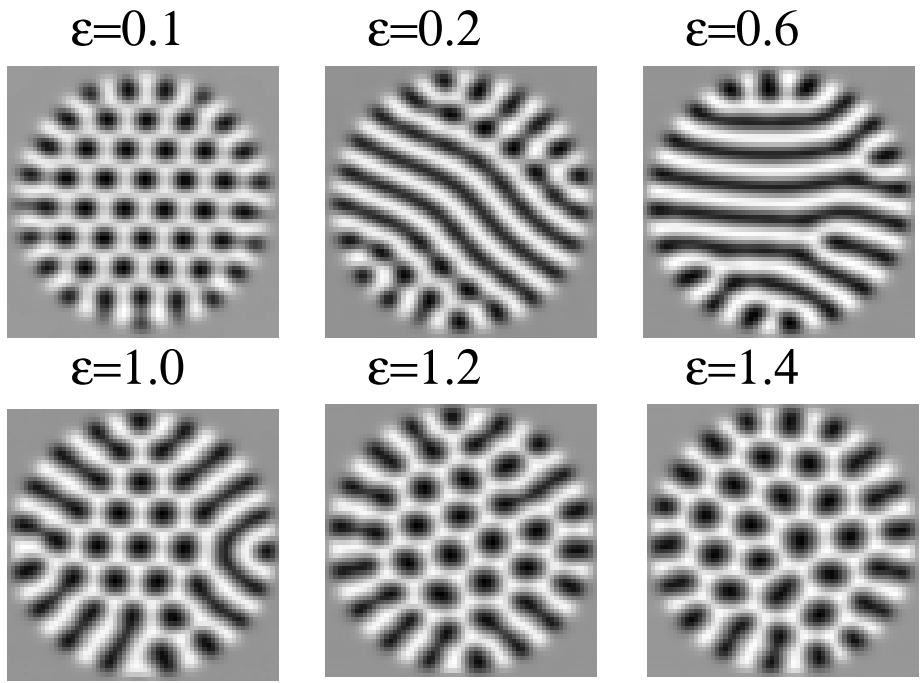}
\caption{Succession of snapshots for $T_0=24\;^oC$ in a circular cell of water 
of thickness $d=1.8\,mm$. The diameter of the cell is $L=8\cdot 2\pi/q_c$ and
the snapshots correspond to an integration time of $400\,t_v$. 
\LB{to24seq}} 
\end{figure}
\end{center}

In order to mimic the experimentally employed circular
containers we apply a strong radial subcritical ramp in the
Rayleigh number that suppresses any convection outside a certain
radius. Such a boundary condition is found to prefer rolls
perpendicular to this `wall' (cf. \cite{DePe94a}, Fig.1 in
\cite{BoPe00}).  Snapshots of simulations in such a circular
cell with diameter $L=8\cdot 2\,\pi/q_c$ and layer thickness
$d=1.8\,mm$  are shown in Fig.~\ref{to24seq}. In all cases 
random initial conditions were used. Near threshold
($\epsilon=0.1$) the cell is completely filled with regular
hexagons, except for a narrow ring of roll-like structures that
are driven by the boundaries. When the heating is increased 
($\epsilon=0.2$) rolls in the interior become linearly
amplitude-stable (cf. Fig.\ref{ampli-water}) and the rolls that
are driven by the boundary invade most of the cell. So far the
scenario is very similar to the experimental observations shown
in Fig.4 of \cite{PaPe92} for $T_0=28^oC$ (see also Fig.
\ref{simu-stabi} below). While the roll pattern persists near
the boundaries for yet larger values of $\epsilon$, a domain of
ordered hexagons appears in the center of the cell for
$\epsilon={\mathcal O}(1)$. The size of the inner domain of
hexagons grows with increasing $\epsilon$ and for $\epsilon=1.4$
the hexagons fill essentially the whole convection cell. The
growing of the hexagon domain with increasing $\epsilon$ can be
understood to arise from a balance between the increasing
tendency of the hexagon domains to invade domains of rolls on
the one hand and the predominance of rolls near the boundaries
on the other hand.  It is worth noting that according to our
computations \cite{PaPe92} should have obtained stable reentrant
hexagons if they had gone to larger values of the Rayleigh
number.

\begin{center}
\begin{figure}
\centering
\includegraphics[width=0.8\textwidth]{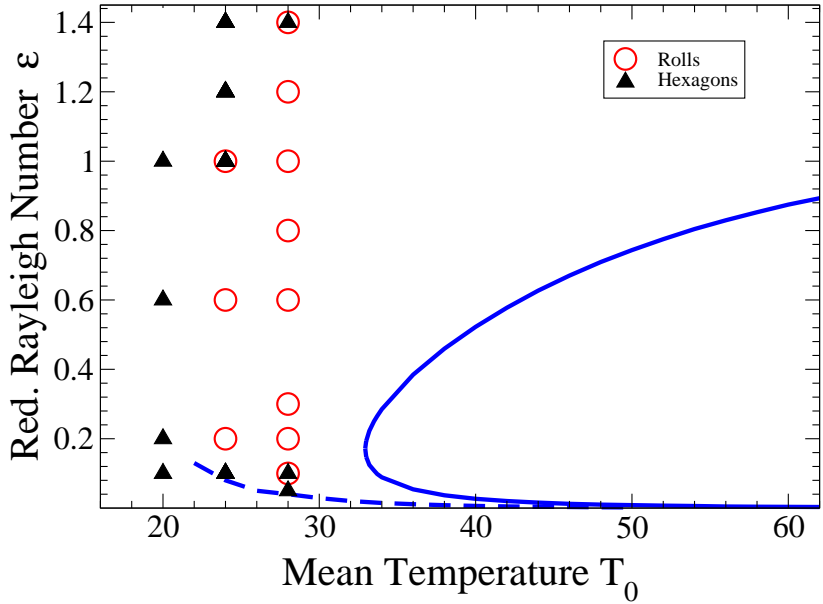}
\caption{Results of  simulations for a circular cell of 
thickness $d=1.8\,mm$ and diameter $L=8\cdot 2\,\pi/q_c$. The simulations
have been carried out for $T_0=20\,^oC,\,24\,^oC,\,28\,^oC$. 
Circles correspond to rolls and triangles to hexagons. The stability 
limits for amplitude instabilities for hexagons (full line) and 
for rolls (dashed line) are reproduced from
Fig.\ref{ampli-water}.
\LB{simu-stabi}} 
\end{figure}
\end{center}

A comparison of the snapshot at $\epsilon=0.1$ with that at
$\epsilon=1.4$ shows that the reentrant hexagons arising from
random initial conditions have a smaller wavenumber than the
hexagons that appear at threshold. This trend is consistent with
the results of the side-band stability calculations (Figs.
\ref{water:buble}, \ref{to28-32}), which show that with
increasing $\epsilon$ the wavenumber range of stable hexagons
moves toward lower wavenumbers. 

In Fig.~\ref{simu-stabi} the results of a set of simulations in
a  circular cell with the same dimensions as in
Fig.~\ref{to24seq} are summarized for three mean temperatures, 
$T_0=20\,^oC$, $T_0=\,24\,^oC$, and $T_0=\,28\,^oC$.  Again
random initial conditions are used. The triangles denote
parameter values for which the final state consists mostly of
regular hexagons, while the circles indicate a final roll state.
When both patterns coexist over the course of the simulation
($t_{max}=200t_v$) both symbols are plotted. The simulations
show that for $T_0=20\,^oC$ the NOB-effects are so strong that
hexagons dominate rolls over the full range of
$\epsilon$ studied. As seen before in Fig.\ref{to24seq}, for a
mean temperature of  $T_0=24\,^oC$  an intermediate range of
$\epsilon$ arises in which the final state consists of rolls. 
For  $\epsilon=1.0$ rolls and hexagons coexist and for yet
larger $\epsilon$ reentrant hexagons appear. If the mean
temperature is increased further to $T_0=28\,^oC$ the same
sequence: hexagons $\rightarrow$ rolls+hexagons
$\rightarrow$ rolls $\rightarrow$ rolls+hexagons
$\rightarrow$ hexagons is obtained. The only difference to the
case $T_0=24^oC$ is that the reentrance is shifted to larger
values of $\epsilon$. 

To make contact with the linear stability analysis for periodic
boundary conditions Fig.~\ref{simu-stabi} shows also the
stability limits for hexagons and rolls with solid and dashed
lines, respectively. For the comparison with the results of the
simulations in the circular container it is important to keep in
mind that the point at which the hexagons and rolls have equal
energy, or more generally, at which the velocity of fronts
separating the two states changes sign is not given by the
linear stability limit of either the hexagons or the rolls, but
rather lies somewhere between the two stability limits. It is
therefore reasonable to expect that as the upper linear
stability limit of hexagons is shifted to lower values of the
Rayleigh number with decreasing mean temperatures $T_0$, the
Rayleigh number at which hexagons start to invade rolls also
decreases and correspondingly the transition from rolls to
hexagons in the presence of boundaries is also shifted to lower
Rayleigh numbers.

\section{Conclusions \LB{sec:conclusions}}

In this paper we have studied non-Boussinesq convection in water
for realistic parameters and boundary conditions. We have
complemented numerical stability analyses for periodic boundary
conditions with  direct numerical simulations that mimic set-ups
used in usual laboratory experiments.  Our main result is the
finding of {\it reentrant} hexagons, i.e. we find that the
hexagon patterns, which typically become unstable not too far
from threshold (\cite{Bu67}), can regain stability further above
threshold via a specific restabilization transition. For strong,
but realistic NOB-effects  hexagons that become unstable at
$\epsilon=0.15$ can become stable again already at
$\epsilon=0.2$. For yet stronger NOB-effects hexagons are
amplitude-stable over the whole range of $\epsilon$ investigated
($\epsilon < 1.5$). This stabilization over the whole range of
$\epsilon$ is not due to a shifting of the initial transition
from hexagons to rolls to ever increasing $\epsilon$, but rather
to the collision of this transition with the restabilization
transition of the hexagons.

Reentrant non-Boussinesq hexagons have been observed in
convection experiments using $SF_6$ near its thermodynamic
critical point as a working fluid ({\cite{RoSt02}). Their
restabilization has been attributed to the strong compressibility
in this regime. Our computations, being based on water as the
working fluid, demonstrate that compressibility is not necessary
for this phenomenon. We show that in water the reentrance is
instead connected with the fact that even in the Boussinesq-case
hexagons can become stable for sufficiently large $\epsilon$
(\cite{AsSt96,ClBu96}) and with the increase of the
non-Boussinesq effects with $\epsilon$. A simple
amplitude-equation model capturing the latter aspect provides
qualitative insight into the stability of the hexagons, including
their side-band instabilities. 

Reflecting the fact that Boussinesq hexagons are stable with
respect to side-band perturbations only for
low wavenumbers (\cite{ClBu96}), we find that in water the
wavenumber of the reentrant hexagons is noticeably below $q_c$.
In contrast, the strongly nonlinear hexagons found experimentally
in $SF_6$ do not exhibit this trend (\cite{RoSt02}). Since we
find the same tendency towards lower wavenumbers also in gases
away from the thermodynamic critical point (\cite{MaRi05a}), one
might speculate that the stabilization of the experimentally observed
hexagons is due to a mechanism that differs from the mechanism of
reentrance discussed here. In the experimental $SF_6$-system even
small changes in the mean density can modify the non-Boussinesq
effects significantly, which would strongly affect the
transitions between rolls and hexagons (\cite{Ah05}). Compared to
$SF_6$ near the critical point, water has the great advantage
that the non-Boussinesq effects are much easier to control.

The connection between the reentrant hexagons and the Boussinesq
hexagons suggests that even in the non-Boussinesq case
down-hexagons may become linearly amplitude-stable, albeit for
yet higher Rayleigh numbers. We show that this is indeed the
case. We have not investigated their side-band instabilities and
it is not clear whether they can coexist with up-hexagons or
whether domains of up-hexagons always invade domains of
down-hexagons. 

To address the stability and dynamics of the hexagon patterns in
large-aspect ratio systems with non-periodic boundary conditions
we have also performed direct numerical simulations of the
Navier-Stokes equations. By a suitable strong variation of the
local Rayleigh number we have implemented a qualitatively
convincing model of a circular container (\cite{DePe94a}). For
intermediate NOB-effects we confirm the experimentally observed
scenario in the low-$\epsilon$ regime (\cite{PaPe92}). Our
computations suggest that even in this experimental set-up
reentrant hexagons should have been accessible experimentally
for somewhat stronger heating ($\epsilon \approx 1.5$).  

Since in experiments the side walls always induce roll
convection the transition between  rolls and hexagons occurs
when the velocity of a front separating the two states changes
sign and not at the stability limit of the hexagons. It is
therefore not possible to study the amplitude-stability limit of
the hexagons in such cells. We expect, however, that by forcing
hexagonal patterns near the walls either by an appropriate
space-dependent heating (\cite{SeSc02}) or by a suitably
corrugated bottom plate (\cite{Bounpub}) hexagons can be
stabilized near the walls and the bulk stability of hexagons
with respect to rolls can be investigated. 

The merging of the lower stability limit of hexagons with the
restabilization line provides also an explanation for the
large contiguous stability range that was found in rotating
non-Boussinesq convection using water (\cite{YoRi03b}). An
interesting question is how the restabilization interacts with
the Hopf bifurcation of the hexagons to oscillating hexagons,
which is induced by rotation (\cite{Sw84,So85,EcRi00a,MaPe04}). 

We thank G. Ahlers for providing us with the code to determine
the non-Boussinesq coefficient and for stimulating discussions.
We gratefully acknowledge support by the office of Basic Energy
Sciences at the Department of Energy (DE-FG02-92ER14303) and an
NSF-SCREMS equipment grant (DMS-0322807).

\bibliography{journal}
\bibliographystyle{jfm}

\end{document}

%% file: jfmheader.tex
\let\realverbatim=\verbatim
\let\realendverbatim=\endverbatim
\renewcommand\verbatim{\par\addvspace{6pt plus 2pt minus 1pt}\realverbatim}
\renewcommand\endverbatim{\realendverbatim\addvspace{6pt plus 2pt minus 1pt}}

\ifCUPmtlplainloaded
\else
\fi
\ifCUPmtlplainloaded

\fi
\ifCUPmtlplainloaded
  \font\bit = mtmib10 at 10.5pt \skewchar\bit ='177  
\else
  \font\bit = cmmib10 \skewchar\bit ='177  
\fi
\newsavebox{\thalfbox}
\sbox{\thalfbox}{$\textstyle\frac{1}{2}$}

\newsavebox{\shalfbox}
\sbox{\shalfbox}{$\scriptstyle\frac{1}{2}$}

\newsavebox{\squartbox}
\sbox{\squartbox}{$\frac{1}{4}$}

\newsavebox{\etbox}
\sbox{\etbox}{\boldmath$\eta$}

\newsavebox{\biSbox}
\sbox{\biSbox}{\bit S}

\newsavebox{\astrutbox}
\sbox{\astrutbox}{\rule[-5pt]{0pt}{20pt}}

\mathchardef\varLambda="0103

\ifCUPmtlplainloaded
\else
\fi
\ifCUPmtlplainloaded
\else
  \font\tenbms=cmbsy10          \skewchar\tenbms ='60
  \font\sevenbms=cmbsy10 at 7pt \skewchar\sevenbms ='60
  \font\fivebms=cmbsy10 at 5pt  \skewchar\fivebms ='60

  \newfam\bmsfam
  \textfont\bmsfam=\tenbms
  \scriptfont\bmsfam=\sevenbms
  \scriptscriptfont\bmsfam=\fivebms

\fi